\documentclass[journal]{IEEEtran}

\usepackage{xcolor,soul,framed} %,caption
\usepackage{newunicodechar}
\newunicodechar{−}{\textminus}

\colorlet{shadecolor}{yellow}
\usepackage[pdftex]{graphicx}
\graphicspath{{../pdf/}{../jpeg/}}
\DeclareGraphicsExtensions{.pdf,.jpeg,.png}

\usepackage[cmex10]{amsmath}
\usepackage{amssymb}
\usepackage{array}
\usepackage{mdwmath}
\usepackage{mdwtab}
\usepackage{eqparbox}
\usepackage{booktabs}
\usepackage{url}
\usepackage{amsfonts}
\usepackage{graphicx}
\usepackage{subcaption}
\usepackage{graphicx}
\usepackage{color}
\usepackage{tikz}
\usepackage{float}

\usetikzlibrary{shapes.geometric, arrows.meta, positioning}
\usepackage{adjustbox}
\bstctlcite{IEEE:BSTcontrol}

\begin{document}
\bstctlcite{IEEEexample:BSTcontrol}
   \title{Hierarchical Agglomerative Clustering for Efficient Annual Voltage Security Assessment in Very-High RES Penetrated Power Systems }
  % Operating Regime Identification and Analysis for Voltage Security Assessment in Very-High Renewable Penetration Systems
  \author{Rock~Agon,~\IEEEmembership{Student Member,~IEEE,} Robin~Preece,~\IEEEmembership{Senior Member,~IEEE,} Jovica V.~Milanović,~\IEEEmembership{Fellow,~IEEE}

}  
% ====================================================================
\maketitle
\begin{abstract}
Voltage security assessment in power systems with high renewable energy source (RES) penetration requires analyzing many operating conditions to capture variability and uncertainty, but simulating a full year of operating points is computationally costly—motivating the selection of representative operating points (ROPs). Most existing methods cluster demand and generation profiles, but similarity in these profiles does not guarantee similarity in voltage behavior, since reactive power limits, voltage-control actions, and nonlinear network interactions shape voltage response in ways that cannot be inferred from power profile patterns. This paper proposes an unsupervised learning framework that selects ROPs based on the system's actual voltage response: each operating point is represented by system-wide voltage-risk indices from AC power-flow solutions, Principal Component Analysis reduces dimensionality, and Hierarchical Agglomerative Clustering with Ward linkage identifies representative voltage regimes. A comprehensive set of evaluation criteria then measures how well the selected ROPs reproduce the full year's voltage-security characteristics under normal and contingency conditions. On the IEEE Voltage Test System under very high RES penetration, the framework reduces the annual operating point set by 99.66\% while reproducing full-year voltage behavior with 98.3\% reconstruction accuracy in steady state and 93.4\% in post-contingency response, outperforming existing injection-space clustering and heuristic sampling.

\end{abstract}
% === KEYWORDS ====================================================================
% =================================================================================

\IEEEpeerreviewmaketitle

\section{Introduction}

The transition toward clean energy systems is essential for decarbonization, but it introduces significant operational challenges. High RES penetration increases stochastic variability, reduces synchronous inertia, and alters reactive power availability, fundamentally changing stability dynamics \cite{ref3}, \cite{ref4}. Among the various stability concerns, voltage stability has become one of the most critical challenges in modern power systems \cite{ref5}. Governed by a complex interplay between fluctuating generation, load characteristics, network topology, and reactive power capabilities \cite{ref6}, voltage instability threatens secure system operation. These challenges become even more pronounced in systems with very high renewable penetration, where inverter-based resources increasingly displace synchronous generators. The resulting reduction in system strength and the prevalence of power-electronic controls introduce additional nonlinearities and control interactions that significantly influence voltage behavior \cite{ref3}.

To address these uncertainties, risk-based voltage security assessment frameworks have been developed, where security is quantified as the expected impact of adverse voltage events under uncertainty \cite{ref8, ref10}. In planning studies, probabilistic and sequential AC simulations over annual horizons are commonly employed to capture the system dynamics \cite{ref13}. However, full-year Monte Carlo or sequential simulations are computationally costly \cite{ref14}. Scenario reduction and ROPs selection are therefore widely used to approximate the full stochastic operating space using a reduced subset of representative conditions \cite{ref15}.

Traditionally, ROPs are selected using heuristic approaches, where a few typical days (e.g., peak summer, peak winter, light load) are chosen based on engineering judgment \cite{ref16}. The number of scenarios is often fixed a priori and derived from coarse temporal stratification rather than intrinsic system structure. As demonstrated in \cite{ref17}, this judgment-based approach becomes inadequate in systems with high RES penetration where the operating space expands from a few natural patterns to many diverse operating modes. These observations become particularly pronounced in very high RES penetration systems, where inverter-based resources introduce additional nonlinearities with respect to voltage regulation. As a result, voltage regimes no longer correlate strongly with typical RES or demand profiles, making the intuition behind selecting “typical days” unreliable. 

In parallel, clustering-based techniques have been introduced as a more rigorous alternative for selecting ROPs, providing structured coverage and enhanced representativeness of the operating space. For example, in \cite{ref15}, hierarchical clustering has been used to identify ROPs for capacity expansion modeling, while the authors of \cite{ref17} used k-means++ for operation-mode analysis, demonstrating that traditional seasonal representations become ineffective under high RES penetration. However, most existing clustering techniques rely primarily on statistical similarity of load and renewable generation profiles \cite{ref18}-\cite{ref22}. By clustering operating conditions based on demand and generation patterns (referred to as the \textit{injection space} in this paper), these approaches implicitly assume that operating points (OPs) with similar power injection profiles will exhibit similar voltage behavior. In practice, however, similarity in demand and generation patterns does not necessarily imply similar voltage-security behavior. Voltage responses are strongly influenced by network characteristics and control mechanisms, including reactive power limits, control-mode switching of generators and power-electronic resources, and variations in grid strength. These factors interact through the nonlinear AC power flow, shaping the system’s voltage profile in ways that cannot be inferred solely from power injection patterns. Consequently, two OPs with similar net injections may produce significantly different voltage distributions. Conversely, OPs with distinct injection patterns may yield comparable voltage responses when network constraints dominate system behavior. As a result, clustering based solely on demand and generation profiles may fail to preserve voltage-security characteristics. 

These observations motivate a structural shift from injection space clustering to clustering of the system response (referred to as \textit{voltage-response space clustering} in this paper). Moreover, prior works rarely incorporate reactive power reserve margins associated with overvoltage mitigation, despite the increasing frequency of overvoltage phenomena in renewable-dominated systems. This paper proposes a hierarchical agglomerative clustering (HAC) framework for very-high RES-dominated power systems, where operating hours are structured directly in the nonlinear AC voltage-response space using physics-guided voltage-risk indicators. Beyond proving the robustness of voltage-response-space clustering, it is shown that HAC with Ward linkage is the most effective technique for this framework. 

The main contributions of this paper are as follows:
\begin{itemize}

\item Demonstration of the limitations of injection-space clustering for voltage-security assessment, revealing that similarity in net active-power profiles does not result in network voltage response similarity, particularly under contingency conditions.

\item Proposal of a voltage-response-space hierarchical clustering framework for representative operating point selection based on the system's actual voltage response. This ensures that the selected ROPs represent the system voltage regimes with an accuracy of 98.3\% in steady-state and 93.4\% in contingency analysis while compressing the scenario set by 99.66\%.

\item  Development of a comprehensive set of evaluation criteria for representative operating points, including statistical and quantile similarity, distribution-shape similarity, extreme tail-risk coverage, and post-contingency accuracy.
\end{itemize}

% === II. Harmonically-Terminated Power Rectifier Analysis=========================================================================================================

\section{Methodology}
The main purpose of this work is to select ROPs based on the voltage response of the system rather than its power patterns. Instead of clustering demand and generation profiles, OPs are represented using risk-based indices derived from AC power-flow solutions. Clustering is then performed in this voltage-response feature space to identify groups of OPs that exhibit similar voltage-security characteristics. From each cluster, an ROP is selected and evaluated to ensure that the reduced dataset preserves both normal and contingency voltage behavior. The approach consists of four stages: (i) generation of a year-long operating dataset under high-RES conditions, (ii) construction of voltage-response features that characterize system-wide voltage risk, (iii) dimensionality reduction to remove redundancy among these features, and (iv) clustering of OPs to identify representative voltage operating regimes. 
\subsection{Stage I: Operating Points Generation}

The contribution of this work is the \emph{framework} for ROPs selection. Real-world or synthetic data could be used depending on availability and period of study. A synthetic approach is adopted here as the study targets a specific instantaneous RES penetration level. This provides control over the penetration level and operating conditions under study. The distribution models employed are well-established and realistic.

A chronological stochastic simulation framework is adopted to generate a full-year voltage-response dataset under a very high RES penetration. Load demand is modeled using deterministic seasonal and diurnal profiles scaled by a stochastic multiplicative factor. The stochastic component follows a Gaussian distribution to represent random deviations around the expected hourly load level. For PV generation, a deterministic solar irradiance profile capturing diurnal and seasonal patterns is combined with a Beta-distributed random variable representing short-term variability caused by cloud cover. Wind generation is derived from wind speeds sampled from a Weibull distribution and converted to electrical power through turbine power curves. To capture spatial dependence among RES plants, a Student-t copula is used to model correlations between geographically distributed RES units and preserve joint extreme events. Finally, temporal persistence is introduced through a first-order autoregressive AR(1) process applied to the PV and wind generation time series to reproduce realistic temporal correlations.

For each hourly data point, a DC unit commitment is solved to obtain an economically feasible dispatch, followed by AC power flow to compute system states. Simulations are performed sequentially so that the system state at hour t initializes the simulation at hour t+1. In particular, generator dispatch levels, commitment status, controller setpoints, and reactive power outputs are carried forward between time steps. This preserves realistic operating continuity and ensures that reactive limits and controller responses evolve consistently over time. 

\subsection{Stage II: Voltage-response Space: Feature Representation}

Voltage security refers to the ability of a power system to maintain acceptable voltage levels at all buses under normal and disturbed operating conditions \cite{ref23}. Therefore, it is considered legitimate to characterize the voltage condition of the system through system-wide indicators rather than analyzing individual bus voltages separately. Therefore, a set of seven system-wide voltage-risk indices is constructed to summarize the main characteristics of the voltage profile at each OP. The proposed framework characterizes operating points based on their actual voltage response obtained from AC power flow simulations, rather than local sensitivity measures. Therefore, these defined indices capture key aspects of voltage behavior, including extrema, spatial dispersion, violation magnitude and extent, and reactive power margins. Together, they provide a voltage-risk vector describing the system’s nonlinear voltage response and define the proposed voltage-response space used for clustering.
\\

Let $V_i(t)$ denote the voltage at bus $i \in \{1,\dots,N_b\}$, $V^{\text{low}}=0.95$ and $V^{\text{high}}=1.05$ p.u. The indices are as follows: 

\subsubsection{Voltage extrema} 
This index captures the lowest and highest voltage levels in the network, indicating potential undervoltage or overvoltage conditions.
\begin{equation}
V_{\min} = \min_i V_i, 
\qquad
V_{\max} = \max_i V_i.
\end{equation}

\subsubsection{Voltage dispersion} 
This index measures how widely voltages are spread across the system.

For $\bar V$, the mean voltage across all buses,
\begin{equation}
\text{Std} =
\sqrt{
\frac{1}{N_b}
\sum_{i=1}^{N_b}
\left( V_i - \bar V \right)^2
}
\end{equation}

\subsubsection{Violation severity}
This index quantifies the magnitude of voltage violation at each bus. 
\begin{equation}
\delta_i =
\max(0, V^{\text{low}} - V_i)
+
\max(0, V_i - V^{\text{high}}).
\end{equation}
The RMS violation severity is
\begin{equation}
\text{RMS\_VVS} =
\sqrt{
\frac{1}{N_b}
\sum_{i=1}^{N_b}
\delta_i^2
}.
\end{equation}
\subsubsection{Violation spatial extent}
This index measures the proportion of buses experiencing voltage violations.
\begin{equation}
P_{\text{bad}} =
\frac{1}{N_b}
\sum_{i=1}^{N_b}
\mathbf{1}\!\left(
V_i < V^{\text{low}}
\ \text{or}\
V_i > V^{\text{high}}
\right)
\end{equation}

\subsubsection{Reactive power reserves}
This index quantifies the remaining reactive capability of generators. 
Let $\mathcal{G}$ denote the set of generators, $Q_i$ the reactive power output of generator $i$, and $Q_i^{\max}$ and $Q_i^{\min}$ its reactive power limits.
\begin{equation}
\text{Qres\_Un} =
\sum_{i\in\mathcal{G}}
\left|Q_i^{\max} - Q_i\right|,\quad
\text{Qres\_Ov} =
\sum_{i\in\mathcal{G}}
\left|Q_i - Q_i^{\min}\right|
\end{equation}
\subsection{Stage III: Dimensionality Reduction}
The voltage-risk indices describe different aspects of system voltage behavior, but some of them may be correlated.  Such correlations introduce redundancy and can distort distance calculations used in clustering, as Euclidean distance may give excessive weight to groups of correlated variables. Therefore, a dimensionality-reduction step is introduced to obtain a compact representation of the system voltage-response space.
\\
\subsubsection{Correlation Analysis}
As an initial diagnostic step, the Pearson correlation matrix is computed to identify dependencies between the indices:

\begin{equation}
R_{ij} = \frac{\text{Cov}(X_i, X_j)}{\sigma_i \sigma_j}.
\end{equation}

Strong correlations indicate that multiple indices may carry the same information, justifying the need for dimensionality reduction to improve the stability of distance-based clustering.
\\
\subsubsection{Principal Component Analysis}
To construct a compact and decorrelated feature space, Principal Component Analysis (PCA) is applied. Compared to nonlinear techniques, PCA provides a linear transformation that converts correlated variables into orthogonal components while preserving the dominant variance of the initial voltage-response data.

Let $X \in \mathbb{R}^{T \times m}$ denote the dataset containing $T$ OPs, and $m$ indices and $\tilde{X}$ the centered dataset obtained by subtracting the mean of each feature.

The covariance matrix $\Sigma $ is defined as:
\begin{equation}
\Sigma = \frac{1}{T-1}\tilde{X}^T \tilde{X},
\end{equation}
Eigen-decomposition of the covariance matrix yields:

\begin{equation}
\Sigma = Q \Lambda Q^T,
\end{equation}
where $Q$ is the matrix of eigenvectors and $\Lambda$ is the diagonal matrix of eigenvalues.

Let $Z \in \mathbb{R}^{T \times k}$ denote the resulting PCA-transformed dataset. Projection onto the first $k$ principal components is given by:

\begin{equation}
Z = \tilde{X}Q_k,
\end{equation}

where $Q_k$ contains the first $k$ eigenvectors associated with the largest eigenvalues.

The number of retained components $k$ is selected such that

\begin{equation}
\frac{\sum_{i=1}^{k}\lambda_i}{\sum_{i=1}^{m}\lambda_i} \ge 0.99.
\end{equation}

This criterion ensures that at least 99\% of the total variance of the original feature space is preserved while reducing redundancy between variables. This transformation produces a lower-dimensional, decorrelated representation of the voltage-response space, improving the robustness of the clustering.
\subsection{Stage IV: Clustering and ROP Selection}
Once OPs are represented in the PCA voltage-response space, clustering is used to identify groups of operating conditions that exhibit similar voltage behavior. Each cluster is a group of operating regimes, representing a set of OPs that lead to comparable voltage-security characteristics. Thereafter, an ROP is selected within each cluster to represent that group of OPs with similar voltage behaviors.   
\\

\subsubsection{Voltage-response Space HAC--Ward Clustering}
HAC is adopted in this framework to merge OPs progressively based on
their pairwise distances in the PCA voltage-response space. The
algorithm proceeds bottom-up: each of the OPs
begins as its own singleton cluster, and at every iteration the two
clusters with the smallest inter-cluster distance are merged. This
repeats until all points belong to a single cluster, producing a
complete merge hierarchy that is naturally represented as a
dendrogram. A partition into any desired number of clusters is then
obtained simply by cutting the dendrogram at the corresponding
height, without re-running the algorithm. This reveals the
hierarchical organization of the data and allows the voltage
operating regimes to be examined at different resolution levels,
exposing the nested structure of the operating space without
requiring the number of clusters to be specified a priori.

HAC uses a linkage criterion to quantify the distance between
clusters. Among the available criteria, Ward linkage is selected in
this paper because it minimizes the increase in within-cluster
variance at each merging step [17]. Unlike other linkage criteria, Ward considers the full membership of each cluster and therefore renders compact, statistically coherent clusters, making it suitable for this work.

At each iteration, two clusters $C_a$ and $C_b$ are merged such that the increase in total within-cluster variance is minimized. The merging cost is defined as:
\begin{equation}
\Delta(C_a, C_b) =
\frac{|C_a||C_b|}{|C_a| + |C_b|}
\| \mu_a - \mu_b \|_2^2,
\end{equation}

where the centroids of the clusters are
\begin{equation}
\mu_a = \frac{1}{|C_a|} \sum_{i \in C_a} Z_i , 
\qquad
\mu_b = \frac{1}{|C_b|} \sum_{i \in C_b} Z_j .
\end{equation}

The Ward linkage can be interpreted as a greedy minimization of the total within-cluster sum of squares (WCSS):
\begin{equation}
\text{WCSS} =
\sum_{a=1}^{K}
\sum_{i \in C_a}
\| Z_i - \mu_a \|_2^2 .
\end{equation}

Minimizing WCSS encourages clusters with low internal variance, which facilitates the identification of coherent voltage operating regimes.
\\
\subsubsection{Representative Operating Point Selection}

Once voltage operating regimes are identified, an ROP is selected from each cluster.  The representative is defined as the OP whose position in the PCA space is closest to the centroid of its cluster, thereby best representing the average voltage-response characteristics of that cluster. For cluster $C_c$, the ROP is defined as:

\begin{equation}
ROP =
\arg\min_{i \in C_c}
\| Z_i - \mu_c \|_2 .
\end{equation}
\section{Clustering Performance Evaluation}
Selecting ROPs is effective if the reduced dataset accurately preserves the voltage-security characteristics of the original system behavior. The following evaluation framework therefore assesses how well the ROPs reproduce the statistical distribution, extreme-risk conditions, and geometric structure of the full dataset, with both pre- and post-contingency OPs.
\subsection{Pre-Contingency Performance Evaluation}

This evaluation checks whether the ROPs reproduce the statistical
properties of the full-year dataset. Each ROP represents its cluster
and is weighted by the number of operating points in that cluster:
the value of each ROP is repeated $|C_c|$ times, producing a
weighted empirical distribution of the same size as the full dataset
($T$ points), against which the full-dataset statistics are
compared. Three aspects are examined: preservation of moments and quantile similarity, similarity of the distribution shape, and coverage of extreme voltage regimes.

\paragraph{Moment and quantile Preservation}
For each voltage-risk index $j$ ($j = 1, \dots, m$), three statistics
are compared between the full dataset and the weighted ROP
distribution: the mean, the standard deviation, and the
Value-at-Risk at level $\alpha = 95\%$, $\text{VaR}_{\alpha,j}$ (the
$\alpha$-quantile of the distribution of index $j$). The absolute
errors are:
\begin{itemize}
    \item Mean error: $|\mu_j^{\text{full}} - \mu_j^{\text{rep}}|$
    \item Standard deviation error: $|\sigma_j^{\text{full}} - \sigma_j^{\text{rep}}|$
    \item VaR error: $|\text{VaR}_{\alpha,j}^{\text{full}} - \text{VaR}_{\alpha,j}^{\text{rep}}|$
\end{itemize}
Each error $e_j$ is normalized by the interquantile scale $S_j$,
defined as the interquartile range of index $j$ in the full dataset:
\begin{equation}
S_j = Q_{99}\left(X_j^{\text{full}}\right) - Q_{1}\left(X_j^{\text{full}}\right).
\end{equation}
This places errors from indices with different units and scales onto
a common, bounded scale. The overall moment and quantile preservation accuracy is
the average across all three statistics and all $m$ indices:
\begin{equation}
\text{Acc}_{Mm} = \frac{1}{m} \sum_{j=1}^{m} \left(1 - \frac{|e_j|}{S_j}\right).
\end{equation}

\paragraph{Distributional Geometry}
While moment and quantile preservation checks specific summary statistics, it does
not guarantee that the full shape of the distribution is preserved.
This is assessed using two complementary distance metrics between the
empirical cumulative distribution functions (CDFs) of the full
dataset, $F_{\mathrm{full}}(x)$, and the weighted ROP distribution,
$F_{\mathrm{rep}}(x)$: the Kolmogorov--Smirnov distance $D_{KS}$,
which measures the largest vertical gap between the two CDFs, and the
first-order Wasserstein distance $W_1$, which measures the average
horizontal gap:
\begin{equation}
D_{KS} = \sup_x \left|F_{\mathrm{full}}(x) - F_{\mathrm{rep}}(x)\right|,
\end{equation}
\begin{equation}
W_1(F_{\mathrm{full}},F_{\mathrm{rep}})
=
\int_{0}^{1}\left|F_{\mathrm{full}}^{-1}(u)-F_{\mathrm{rep}}^{-1}(u)\right|du.
\end{equation}
W1 is normalized using the same
interquantile scale $S_j$ used above,
\begin{equation}
 \hat{W}_{1,j} = \frac{W_{1,j}}{S_j}
\end{equation}
Conceptually, this score combines a shape term and a location term in equal proportion such as: 
\begin{equation}
 \text{error} = 0.5[\text{shape error + location error}] 
\end{equation}
A single geometry-accuracy score for the geometry metric is:
\begin{equation}
\text{Acc}_{G,j} = 1 - \frac{1}{2}\left({D}_{KS,j} + \hat{W}_{1,j}\right).
\end{equation}
The overall distributional-geometry accuracy is the average over all
$m$ indices:
\begin{equation}
\text{Acc}_{Gd} = \frac{1}{m}\sum_{j=1}^{m}\text{Acc}_{G,j}.
\end{equation}

\paragraph{Extreme Tail Coverage}
Moment, quantile, and distribution-shape accuracy can be high even if the ROPs
fail to represent the most severe, safety-critical conditions. This
is checked using two indicators based on the RMS voltage violation
severity index, $\text{RMS\_VVS}$: one measures whether the
\emph{magnitude} of extreme violations is preserved, and the other
measures whether extreme \emph{hours} are represented at all.

\textit{Magnitude of extreme violations:} The Conditional
Value-at-Risk (CVaR) is the expected severity of violations beyond
the $\alpha$-quantile of the $\text{RMS\_VVS}$ distribution:
\begin{equation}
\text{CVaR}_\alpha = \mathbb{E}\left[\text{RMS\_VVS} \mid \text{RMS\_VVS} \ge \text{VaR}_\alpha \right].
\end{equation}
The magnitude-accuracy score compares $\text{CVaR}_\alpha$ between the
full dataset and the weighted ROP distribution, normalized by the
interquantile scale of $\text{RMS\_VVS}$:
\begin{equation}
\text{Acc}_{CVaR} = 1 - \frac{|\text{CVaR}_\alpha^{\text{full}} - \text{CVaR}_\alpha^{\text{rep}}|}{S_{\text{RMS\_VVS}}}.
\end{equation}

\textit{Representation of extreme operating states:} The Tail
Coverage Ratio (TCR) measures the fraction of extreme hours in the
full dataset that are correctly represented by an extreme ROP. Let
$\mathcal{T}$ denote the set of extreme hours in the full dataset,
using the same threshold $\text{VaR}_\alpha$ defined above:
\begin{equation}
\mathcal{T} = \{t : \text{RMS\_VVS}(t) \ge \text{VaR}_\alpha\}.
\end{equation}
Let $c(t)$ denote the cluster containing hour $t$, and $r_{c(t)}$ its
representative operating point. TCR is the proportion of extreme
hours whose representative is itself extreme:
\begin{equation}
\mathrm{TCR} =
\frac{
\sum_{t \in \mathcal{T}} \mathbf{1}\left(r_{c(t)} \in \mathcal{T}\right)
}{
|\mathcal{T}|
},
\end{equation}
where $\mathbf{1}(\cdot)$ is an indicator function equal to 1 if the
condition holds and 0 otherwise.

Since $\text{Acc}_{CVaR}$ and $\mathrm{TCR}$ are both defined only for
$\text{RMS\_VVS}$, the overall tail-risk accuracy is their average:
\begin{equation}
\text{Acc}_{\text{TailRisk}} = \frac{1}{2}\left(\text{Acc}_{CVaR} + \mathrm{TCR}\right).
\end{equation}

\paragraph{Reconstruction Score}
The overall pre-contingency reconstruction accuracy combines the
three components above, moment and quantile preservation, distributional
geometry, and tail-risk coverage, with equal weight:
\begin{equation}
\text{Acc}_{\text{Recon}} = \frac{1}{3}\left(\text{Acc}_{Mm} + \text{Acc}_{Gd} + \text{Acc}_{\text{TailRisk}}\right).
\end{equation}

\subsection{Post-Contingency Performance Evaluation}
To evaluate the application of the proposed framework under disturbances, the ROPs are used for contingency analysis. The most critical line contingency is applied in this work, and the resulting voltage responses are compared with those obtained from full-year simulations using Voltage Security Performance Index (VSPI), derived from the commonly-used Performance Index (PI) \cite{ref026}, \cite{ref027}. It aggregates voltage violations across all buses, capturing both the severity of voltage deviations and their spatial extent. Let $V_{\text{low}}$ and $V_{\text{high}}$ denote the lower and upper acceptable voltage limits. The VSPI is then computed:

\begin{equation}
\text{VSPI} =
\sum_{i=1}^{N_b}
\left(
\frac{V_i - V_i^{\text{lim}}}{|1 - V_i^{\text{lim}}|}
\right)^2,
\quad
V_i^{\text{lim}} =
\begin{cases}
V_{\text{low}}, & \text{if } V_i < V_{\text{low}}\\
V_{\text{high}}, & \text{if } V_i > V_{\text{high}}
\end{cases}
\end{equation}

\section{Comparative Analysis: Injection space and heuristic sampling}
To validate the proposed framework, the ROPs selected through voltage-response space clustering have been compared to those selected from the injection space clustering and heuristic sampling, the industry practice.

\subsection{Injection Space: Feature Representation}
The main difference with voltage-response space clustering is that clustering in the injection space is performed using the net active power injection at each bus as the feature vector for each hour.

Let $\mathcal{B}_{\text{inj}}$ denote the set of buses that host at least one load or generation unit. For each hour $t$, the net active-power injection at bus $b \in \mathcal{B}_{\text{inj}}$ is defined as
\begin{equation}
P^{\text{net}}_{b}(t)
=
\sum_{g \in \mathcal{G}_b} P_{g}(t)
+
\sum_{r \in \mathcal{R}_b} P_{r}(t)
-
\sum_{\ell \in \mathcal{L}_b} P_{\ell}(t),
\end{equation}
where $\mathcal{G}_b$, $\mathcal{R}_b$, and $\mathcal{L}_b$ denote the sets of conventional generators, RES units, and loads connected to bus $b$, respectively.

The hourly injection space operating vector is then:
\begin{equation}
\mathbf{x}(t)
=
\big[
P^{\text{net}}_{b_1}(t),\;
P^{\text{net}}_{b_2}(t),\;
\dots,\;
P^{\text{net}}_{b_m}(t)
\big]^{\top},
\quad b_i \in \mathcal{B}_{\text{inj}},
\end{equation}

Adding reactive power to the injection space  features can provide a slight improvement in clustering performance. This approach, however, still underperforms relative to voltage-response-space clustering. Furthermore, achieving comparable results with the augmented injection space requires substantially more representative operating points (ROPs) than either standard injection-space or voltage-response-space clustering. Since the number of ROPs affects computational cost and the fairness of comparative evaluation, this requirement introduces a potential bias. Therefore, to ensure a balanced comparison and adhere to common practice in the literature, this study restricts injection-space clustering to active power injections only.

For the clustering, the 74-dimensional vector x is first normalized and passed through PCA for dimension reduction. 
After PCA processing, K-Medoids++ is applied to identify ROPs. 

K-Medoids++ is selected for its robustness to outliers, which are prevalent in very high-renewable scenarios. Such conditions introduce significant variability and extreme operating points in the feature space. Unlike, HAC where early-stage merges are irreversible and can propagate the influence of outliers throughout the clustering structure, K-Medoids++ mitigates this effect by selecting actual data points as cluster centers and minimizing pairwise dissimilarities. This makes it more suitable for injection space representations with high variability and non-uniform distributions. A preliminary investigation with extensive testing and simulation has confirmed that K-Medoids++ is more appropriate than HAC for injection space clustering. 

\subsection{Heuristic Sampling: Industry Benchmark}
As a baseline industry practice, a heuristic representative selection approach is implemented based on characteristic operating conditions in the injection space. Each hourly operating point can be described by a three-dimensional vector:

\begin{equation}
\mathbf{x}_t =
\left[
P_{\text{load}}(t),\;
P_{\text{PV}}(t),\;
P_{\text{wind}}(t)
\right]^{\top}
\end{equation}
where $P_{\text{load}}(t)$ denotes the total system load, and $P_{\text{PV}}(t)$ and $P_{\text{wind}}(t)$ denote the aggregated photovoltaic and wind generation at hour $t$.

To ensure temporal diversity, the set of operating hours is partitioned into coarse categories based on season (winter, spring, summer, autumn), day type (weekday, weekend, holiday), and time of day (morning, peak-load time, night time).
Within each category, representative operating points are selected by identifying the lowest and highest values of each element of the vector $\mathbf{x}_t$. The resulting set of operating points captures a range of extreme and typical system conditions across both demand and renewable generation, while accounting for seasonal and temporal variability. The total number of selected points is fixed to the optimal number of clusters from HAC and K-Medoids++ to enable direct comparison with the clustering-based ROPs.

\section{Case Study and Numerical Results}
This section presents the case study used to validate the proposed voltage-response-based clustering framework. It includes the description of the test system, the modeling, the generation of the operating data set, and the construction of the voltage-response representation. In addition, the performance of the proposed approach is comprehensively evaluated and compared with injection space clustering and heuristic sampling, under both normal and contingency conditions.
\subsection{Test System Description and Experimental Setup}
\begin{figure}[!t]
  \begin{center}
  \includegraphics[width=\columnwidth]{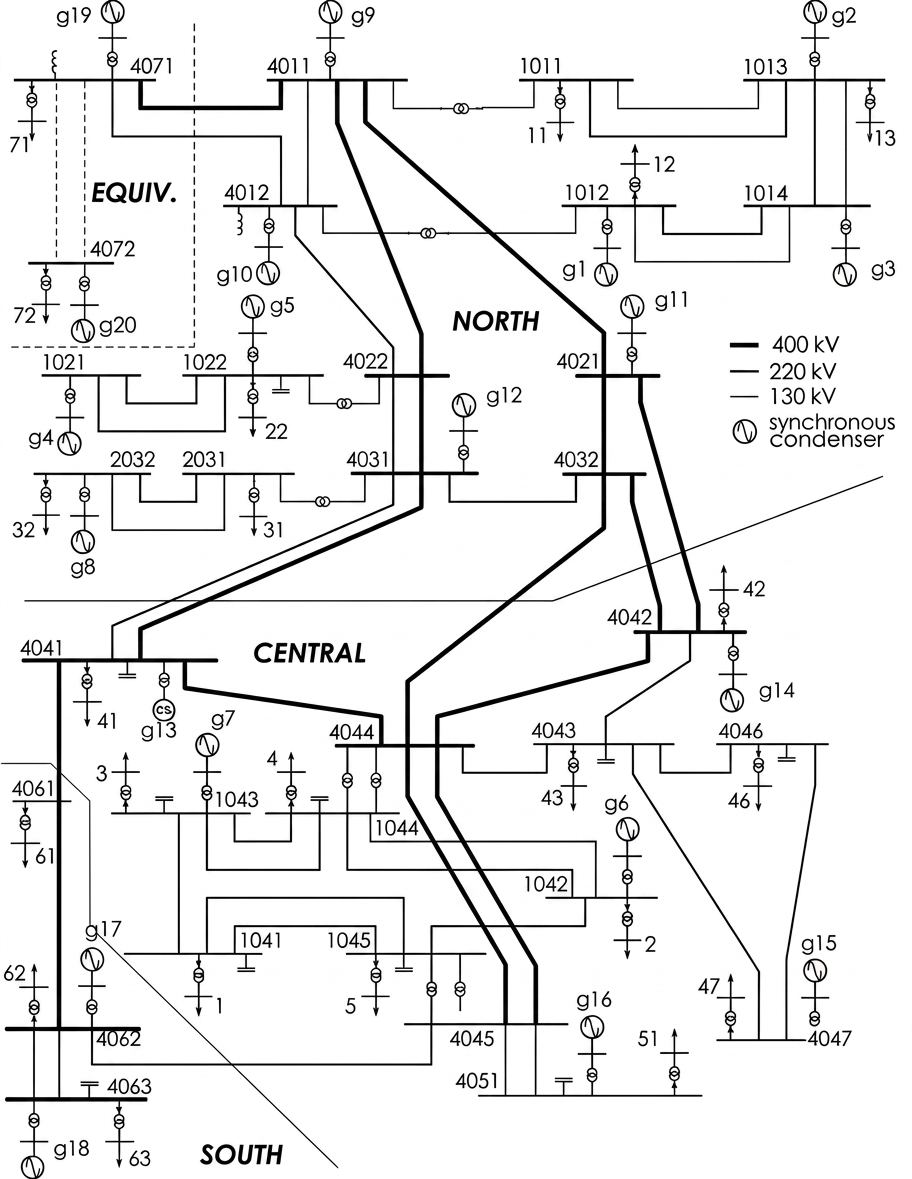}
  \caption{IEEE Voltage Test System}
  \label{f1}
  \end{center}
\end{figure}

The framework is evaluated in DIgSILENT PowerFactory on the IEEE Voltage Test System shown in Fig.~\ref{f1} \cite{ref26}. The system has
been adapted to emulate very high RES penetration conditions. Specifically, central area loads are replaced with WECC composite load models + DER\_A model to capture near-realistic voltage-dependent behavior \cite{ref27}. Wind plants are placed on buses of g6, g7, g14, and g16, while PV plants are placed on buses of g15, g17, and g18. The DERs are PVs behind-the-meter, placed on load buses to cover up to 20\% of local loads. The instantaneous RES penetration level is in the order of 80-90\%. The chronological simulation produces a full‑year dataset of 8{,}760 hourly operating points. %Table I presents the parameters of the stochastic models used to generate the loads, winds, and PV plants dataset.

%\begin{table}[!t]
%\centering
%\caption{Stochastic Model Parameters}
%\label{tab:stochastic_params}

%\begin{tabular}{|l|l|l|}
%\hline
%Component & Model \& Parameter & Value \\
%\hline
%Load  & Normal $(\mu,\sigma^2)$ & $\mu=1.0,\ \sigma=0.05$ \\
%\hline
%PV    & Beta $(\alpha,\beta)$ & $\alpha=1.1,\ \beta=1.7$ \\
%\hline
%Wind  & Weibull $(k,\lambda)$ & $k=2.2,\ \lambda=8.5\,\mathrm{m/s}$ \\
%\hline
%Spatial dep.\ & Student-$t$ copula\ $\nu$ & $\nu=2$ \\
%\hline
%Temporal dep.\ & AR(1), coeff.\ $\phi$ & $\phi_{\mathrm{PV}}=0.85,\ \phi_{\mathrm{wind}}=0.90$ \\
%\hline
%\end{tabular}
%\end{table}

\subsection{Voltage-response Space Representation}
The seven voltage-risk indices have been constructed, and their correlations are shown in Fig. 2. The matrix reveals a level of interdependence among the indices. On the one hand, violation-related metrics are strongly coupled: violations spreading ($P_{\text{bad}}$) and severity metric (RMS\_VVS) exhibit high correlation, while the latter is also strongly linked to the dispersion. Voltage extrema behave asymmetrically. $V_{\max}$ is strongly correlated with the number of buses that violate the limit, whereas $V_{\min}$ shows a weaker association, indicating dominance of overvoltage-driven stress in the studied operating space. On the other hand, reactive reserves exhibit physically consistent relationships with violations: Qres\_Ov is negatively correlated with violation severity, while Qres\_Un is positively correlated. Their perfect anticorrelation confirms that the reactive margin is intrinsically one-dimensional.

\begin{figure}[!t]
    \centering
    \includegraphics[width=\columnwidth]{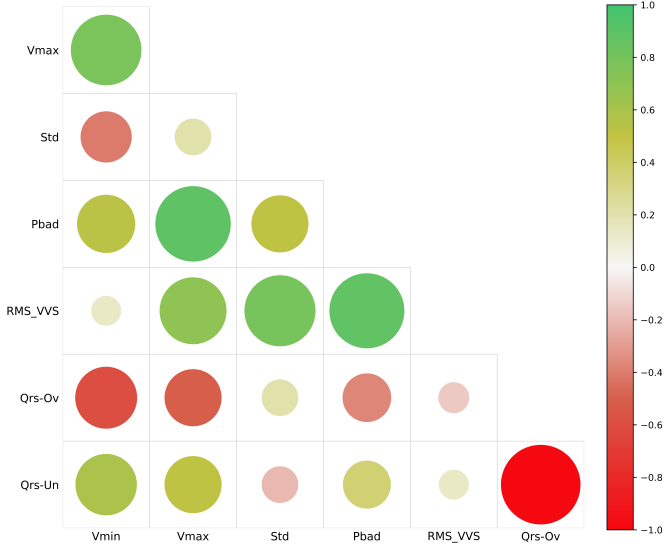}
    \caption{Correlation matrix}
\end{figure}

\subsection{PCA and Geometric Analysis}
PCA is therefore applied to the standardized voltage-risk indices to reduce the dimensionality. 
The scree plot in Fig. 3 confirms the compressibility of the space. PC1 explains 54.13\% of total variance, PC2 explains 33.04\%, and PC3 11.26\%. Achieving a 99\% preservation threshold requires the four components, yielding 99.24\% cumulative variance. 

\begin{figure}[!t]
   \centering
    \includegraphics[width=\columnwidth]{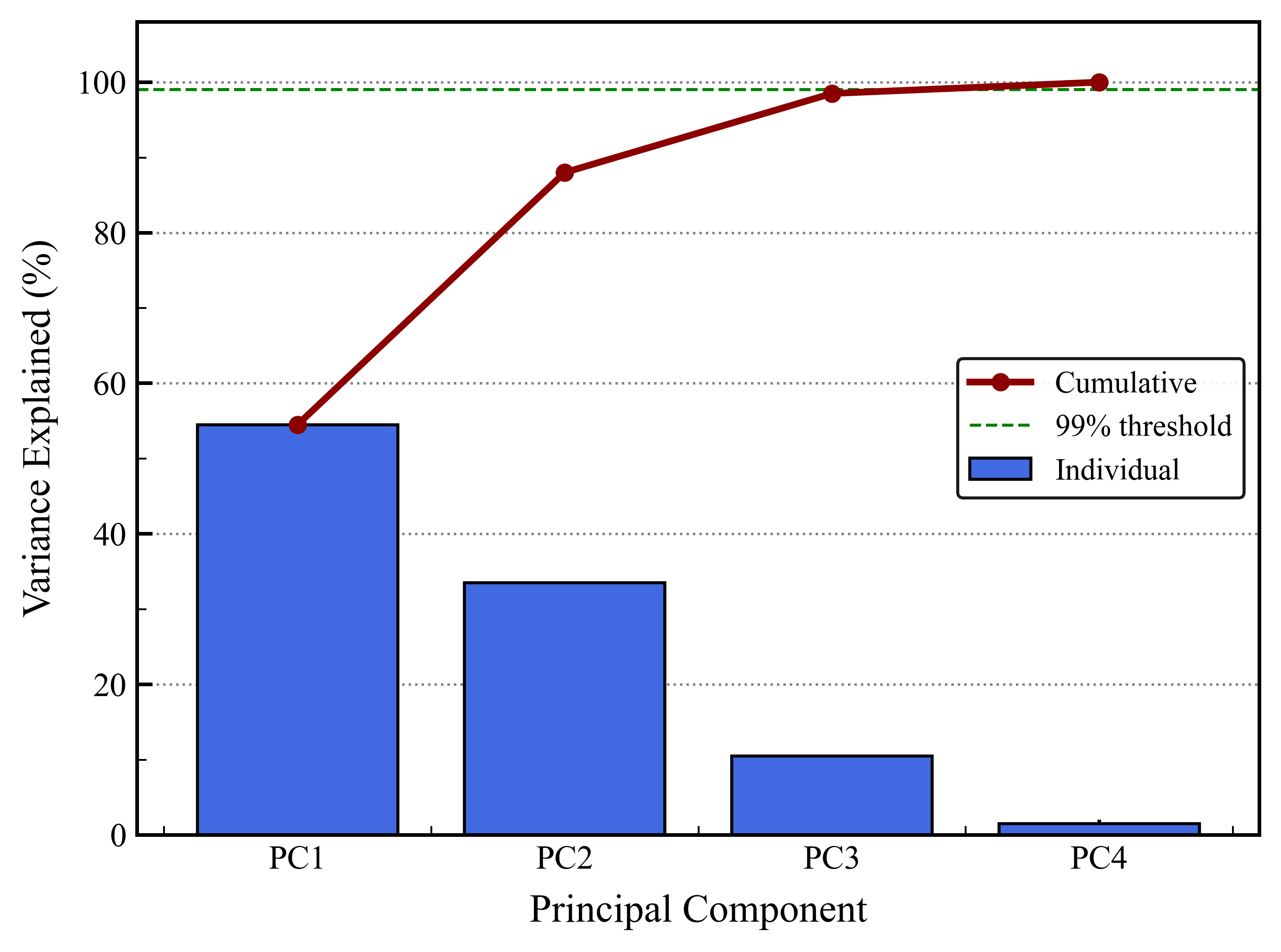}
    \caption{PCA scree plot}
\end{figure}

\subsection{Diagnostic and Clustering}
The elbow method is used to identify the optimal number of clusters for K-Medoids++ to be 30 clusters, as illustrated in Fig. 4.  K-Medoids++ clustering is first applied in the injection space, and the resulting 30 medoids are selected as the ROPs.

For the voltage-response space, HAC does not inherently require a predefined number of clusters. Therefore, a sensitivity analysis is conducted to see if also HAC performs the best with 30 clusters. Fig. 5 shows the reconstruction error as a function of the number of clusters. Improvements beyond 30 clusters are marginal or negative, illustrating that 30 clusters from HAC also performs well. The internal score is the Cophenetic Correlation Coefficient, which measures how faithfully the hierarchical tree, constructed during the HAC process, preserves the pairwise distances between data points.

For benchmarking purposes, a heuristic sampling set of 30 operating points is also constructed. These points are selected to capture representative conditions, including high and low levels in load, PV and wind levels,  across seasons, day/night periods, and weekday/weekend/holiday patterns. The number of samples is fixed to 30 to ensure a fair comparison with the clustering-based ROPs.

\begin{figure}[!t]
  \begin{center}
  \includegraphics[width=\columnwidth]{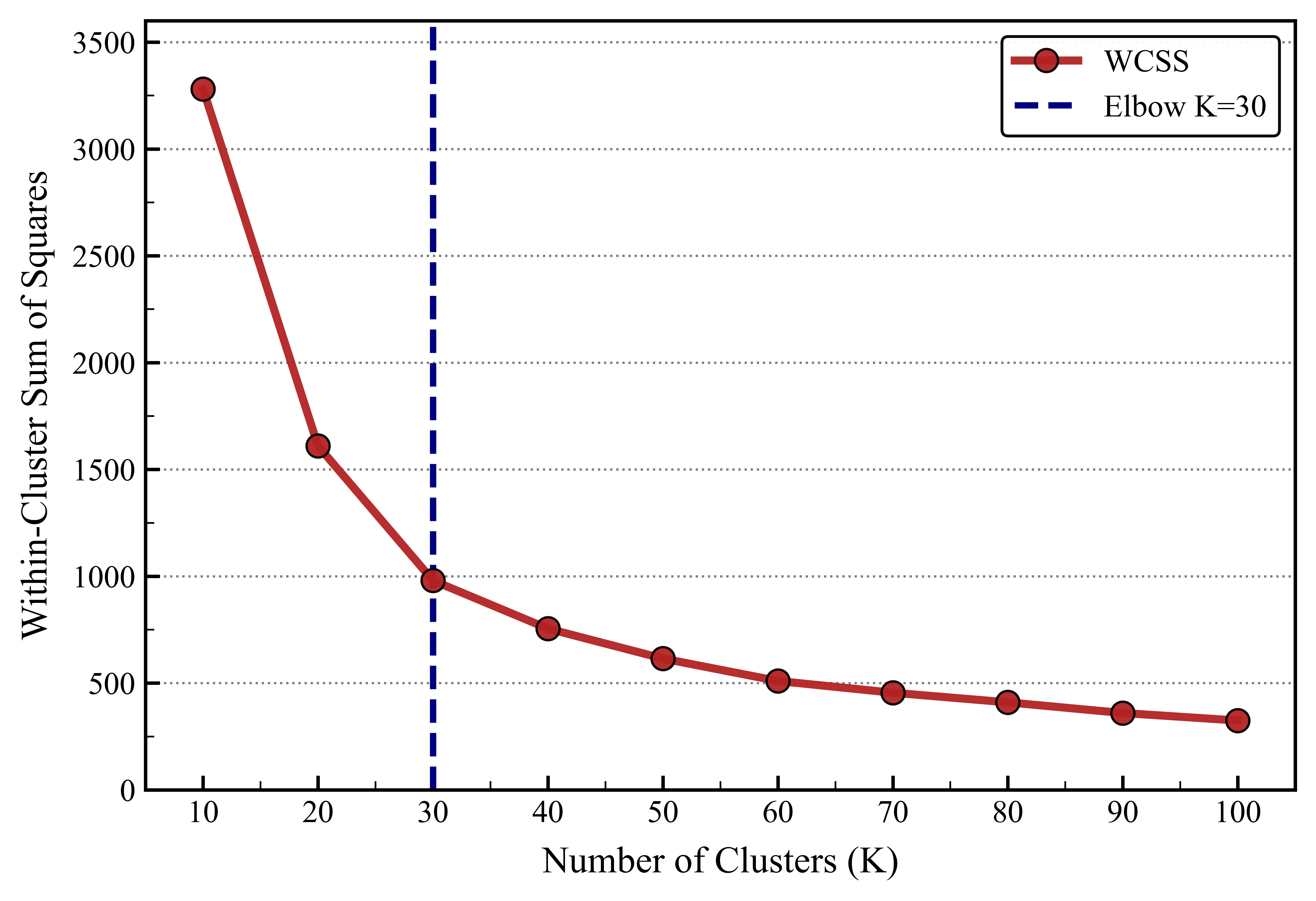}
  \caption{Selection of optimal number of clusters}
  \label{f7}
  \end{center}
\end{figure}

\begin{figure}[!t]
  \begin{center}
  \includegraphics[width=\columnwidth]{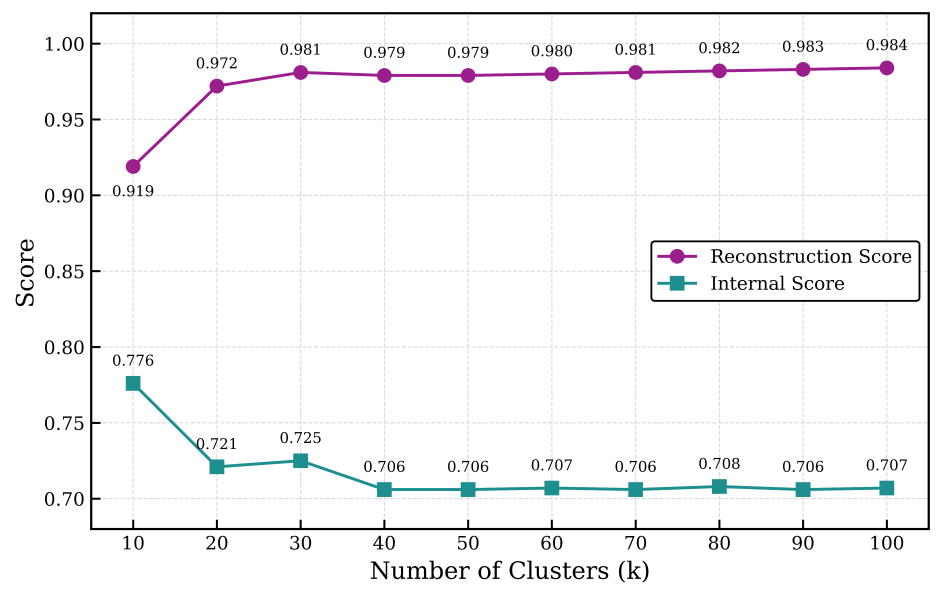}
  \caption{Voltage-response space HAC k-sensitivity performance}
  \label{f8}
  \end{center}
\end{figure}

%\subsection{Clustered Voltage Operating Modes visualization}
%The 2D and 3D t-distributed Stochastic Neighbor Embedding (t-SNE) projections are shown in Fig.~\ref{f14}. The two graphs are meant for visualization, and visual distinction between ROPs groups does not mean physical distances between cluster, but the visualization highlights the local organization of ROPs. However, several structural patterns can be observed. First, a number of clusters appear densely packed and compact, which corresponds to nominal operating conditions characterized by balanced reactive margins and limited voltage stress. In contrast, other clusters form elongated arc-like structures rather than isotropic clouds. This geometry suggests the presence of continuous nonlinear trajectories, likely driven by gradual variations in renewable generation penetration and reactive loading conditions. This spatial organization reinforces the interpretation that voltage operating regimes form a structured continuum of system responses rather than isolated categories.

%----------------------------------------------------
%\begin{figure}[!t]
%\begin{subfigure}{0.5\textwidth}
%  \includegraphics[width=3.5in]{tsne_2D.png}
%  \caption{t-SNE 2D}\label{ciam}
%\end{subfigure}
%\vfill
%\begin{subfigure}{0.5\textwidth}
 % \includegraphics[width=3.5in]{tsne_3D_notitle.png}
 % \caption{t-SNE 3D}\label{ciam}
%\end{subfigure}
%\caption{Cluster visualization in t-SNE}
%\label{f14}
%\end{figure}

\subsection{Pre-Contingency ROPs Evaluation}
\begin{figure}[!t]
  \begin{center}
  \includegraphics[width=\columnwidth]{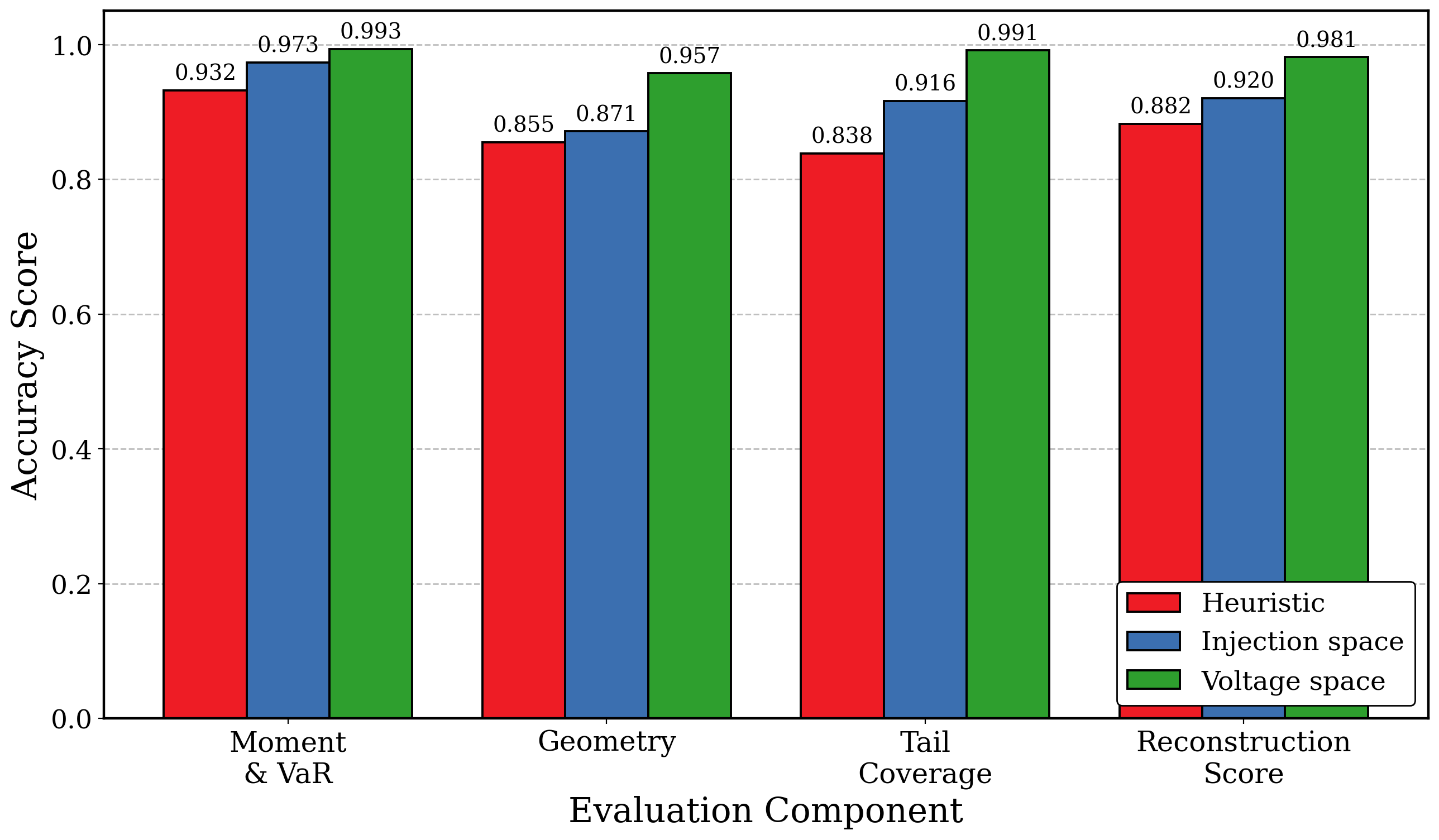}
  \caption{Pre-Contingency performance of 30 ROPs}\label{f9}
  \end{center}
\end{figure}

The objective of this section is to assess whether structuring operating points based on voltage response leads to better preservation of voltage-security characteristics compared to conventional representations. 
The results in Fig.~\ref{f9} demonstrate the superior performance of the proposed framework compared to the injection space clustering and heuristic sampling across all evaluation components. The voltage-response space HAC method consistently achieves the highest accuracy, reaching near-perfect performance in tail coverage and in moment an quantile preservation, while 96.1\% in geometry distribution preservation, indicating its effectiveness in capturing both global statistical properties and critical tail behavior. The injection space approach also shows significant improvement over the heuristic method, particularly in tail coverage, suggesting enhanced representation of extreme operating conditions. In contrast, the heuristic selection yields the lowest performance across all metrics, especially in tail coverage, highlighting its limited ability to capture rare but critical system states. Overall, the reconstruction error of the proposed reduced-voltage-response space HAC is 1.7\% compared to 4.3\% of injection space K-Medoids++ and 12.5\% for heuristic sampling, confirming that clustering-based ROPs, particularly in voltage space, provide a more accurate and comprehensive characterization of system behavior. Therefore, expressed as an overall representativeness score, the voltage‑response space HAC retains 98.3\% of the full‑year steady‑state voltage behaviour.

%-----------------------------------------------------------------------------
\subsection{Post-Contingency ROPs Evaluation}
The comparison is extended to contingency conditions to assess the robustness of the selected ROPs. The objective is to verify whether the reduced set of operating points can reproduce the post-contingency behavior of the full-year dataset with comparable or slightly conservative performance. A slightly conservative representation is generally preferred in security assessment because overestimating system stress is operationally safer than underestimating potentially insecure conditions, although excessive conservatism also lead to unnecessarily high operational costs. 

All single line outages were screened by their effect on the system‑wide VSPI, and line 4032–4044 was retained as the most critical contingency, i.e. the outage producing the largest aggregate voltage violation. The ROPs obtained from each method are evaluated under this contingency and compared against the full‑year simulation.

\begin{figure}[!t]
  \includegraphics[width=\columnwidth]{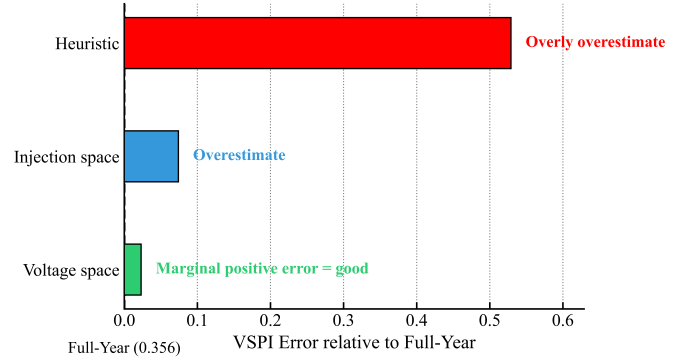}
\caption{Line 4032-4044 post-contingency ROPs performance}
\label{f10}
\end{figure}

 The post-contingency results in Fig.~\ref{f10} highlight substantial differences in accuracy among the three approaches when estimating the VSPI. The heuristic method exhibits a significant overestimation error exceeding 0.5, indicating its limited ability to reliably capture post-contingency system behavior, which would lead to overly conservative planning decisions and potentially higher investment than needed. In contrast, the injection space clustering approach considerably reduces this error to approximately 0.07, demonstrating improved representation of operating conditions, although a noticeable overestimation bias remains. The voltage space method achieves the lowest error, approximately 0.02, with only a marginal positive deviation. This indicates a highly accurate reconstruction (93.4\%) of the full-year VSPI and confirms its ability to capture critical post-contingency voltage dynamics. The small positive error is expected for conservative security analysis purposes. 

The robustness of the proposed method is further examined by removing the system’s only synchronous condenser. As shown in Fig. 8, there is a consistent trend of underestimation across the heuristic and injection space approaches when evaluating the VSPI relative to the full-year reference. The heuristic method exhibits the largest negative bias, indicating a substantial underestimation of system voltage deviation and an inability to capture critical operating conditions. The injection space method reduces this bias but still underestimates the VSPI, suggesting incomplete representation of voltage-critical scenarios. In contrast, the voltage space approach achieves near-zero error with a slight positive deviation, indicating a well-balanced and accurate approximation. This marginal bias is desirable, as it avoids the risk of underestimating system vulnerability. Overall, the results demonstrate that clustering in voltage-response space provides the most reliable approximation and appropriate conservative estimation of voltage stability performance.

\begin{figure}[!t]
  \includegraphics[width=\columnwidth]{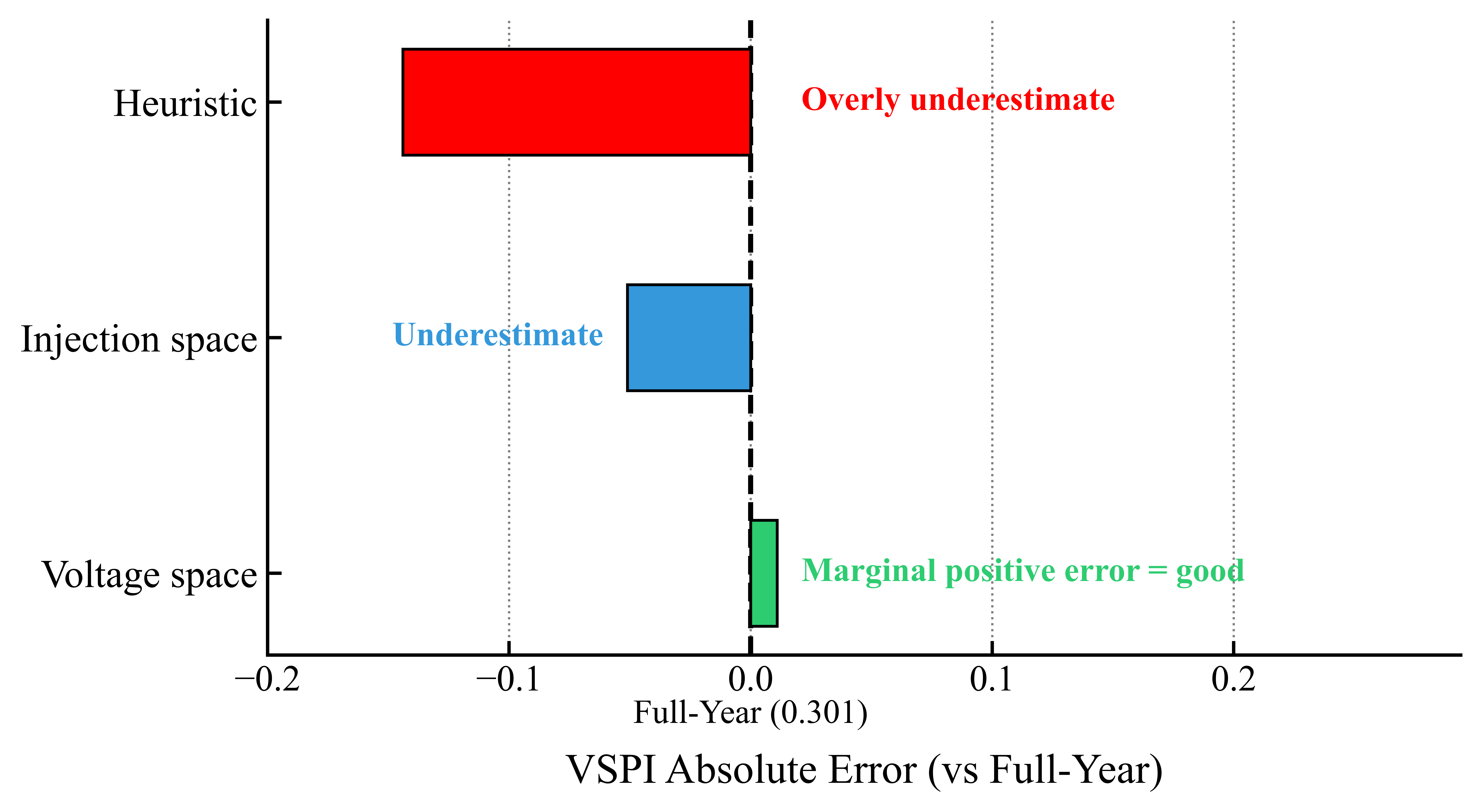}
\caption{SynCondenser post-contingency ROPs performance}
\label{f100}
\end{figure}

%----------------------------------------------------
%\begin{figure}[H]
%  \begin{center}
%  \includegraphics[width=3.6in]{summary_bar (10).png}\\
%  \caption{ROPs Performance Relative Error to Full OPs}\label{f12}
%  \end{center}
%\end{figure} 
%----------------------------------------------------
%----------------------------------------------------
%\begin{figure}[H]
%  \begin{center}
%  \includegraphics[width=3.5in]{evaluation_ward_pre_vls.png}\\
%  \caption{HAC-ward Clustering in voltage-response space}\label{f13}
%  \end{center}
%\end{figure} 
%----------------------------------------------------

%----------------------------------------------

%-----------------------------------------------------------------------

\section{Discussion}
The hierarchical clustering results indicate that voltage behavior in very-high RES is inherently structured. Operating conditions form nested regimes of increasing stress severity. This aligns with voltage physics, where deviations propagate collectively and cluster around shared stress patterns \cite{ref33}-\cite{ref36}. Under normal operating conditions, all methods achieve acceptable performance. However, significant differences emerge under disturbances. Injection-space clustering and heuristic sampling fail to capture critical regimes, leading to large over- or underestimation of voltage risk. In contrast, the proposed voltage-response clustering remains accurate, with only a marginal conservative bias. This demonstrates that preserving statistical properties alone is insufficient; capturing the structure of voltage regimes is essential for reliable disturbance assessment. 

A potential concern is whether the superior performance of the
proposed method stems from the voltage-response feature space or
from the choice of clustering algorithm. To ensure a fair
comparison, each feature space is paired with its best-performing
algorithm: K-Medoids++ for the injection space, which outperforms
HAC there owing to its robustness to the outliers prevalent in that
representation, and HAC--Ward for the voltage-response space. The
proposed method is thus compared against the injection space at its
strongest configuration, rather than a deliberately weakened
baseline. Under these conditions the voltage-response space still
yields substantially lower reconstruction error.

\section{Conclusion}

This paper proposed a voltage-response space clustering framework for identifying representative operating regimes in high-renewable power systems. Results show that while conventional methods perform adequately under normal conditions, only the proposed approach remains accurate and reliable under disturbances. By clustering based on voltage behavior rather than injections, the method preserves critical stability characteristics while reducing computational effort. The framework provides both an efficient representation of operating conditions and a physically meaningful characterization of voltage regimes, making it well-suited for voltage security assessment in renewable-dominated grids.
Future work includes exploring this clustering framework to reduce the continuation power flow runs in data-driven voltage stability constrained power system planning. 

\ifCLASSOPTIONcaptionsoff
  \newpage
\fi


%\begin{biography}[{\includegraphics[width=1in,height=1.25in,clip,keepaspectratio]{mshell}}]{Michael Shell}

% ==== SWITCH OFF the BIO for submission
% ==== SWITCH OFF the BIO for submission
\begin{IEEEbiography} [{\includegraphics[width=1in,height=1.25in,clip,keepaspectratio]{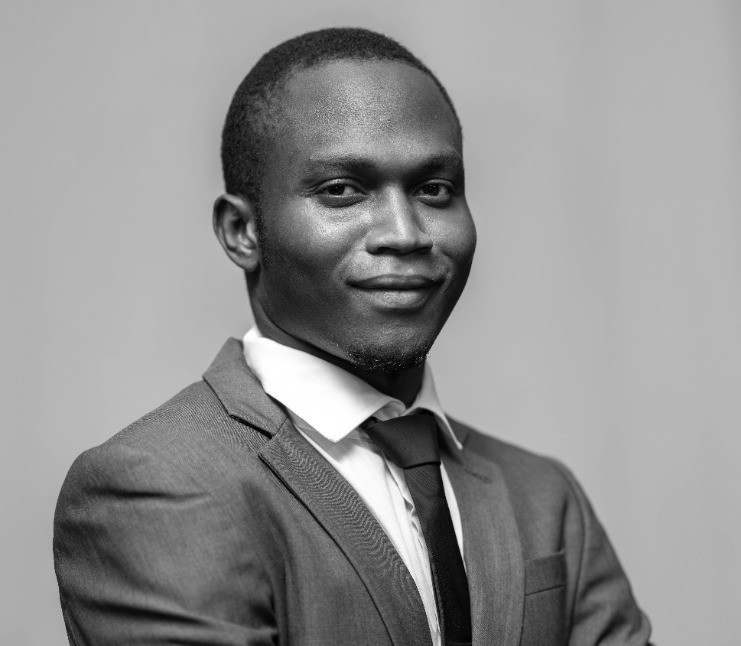}}] {Rock Agon} received the MSc degree in electrical and computer engineering from Carnegie Mellon University, Rwanda, in 2024. He is currently working toward a dual-award Ph.D. degree at The University of Manchester, UK, and Tsinghua University, China. His research interests include power system stability, planning, and optimization with very high renewable energy penetration.
\end{IEEEbiography}
\begin{IEEEbiography}[{\includegraphics[width=1in,height=1.25in,clip,keepaspectratio]{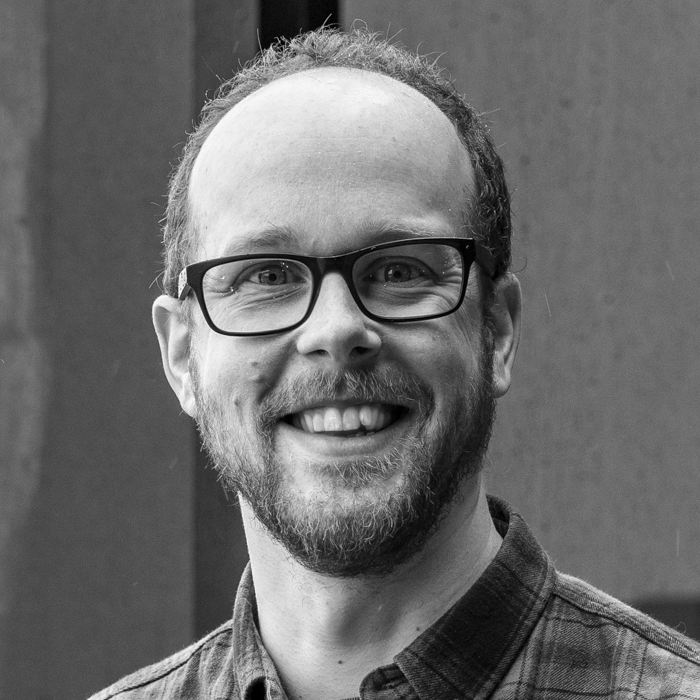}}] {Robin Preece} received the B.Eng. and Ph.D. degrees from The University of Manchester, U.K. He is currently a Professor and Head of Power Systems in the Department of Electrical and Electronic Engineering at the same university. His research interests include power system stability assessment, cascading failures, resilience of low-carbon grids, HVDC modelling, and uncertainty quantification. 
\end{IEEEbiography}

\begin{IEEEbiography} [{\includegraphics[width=1in,height=1.25in,clip,keepaspectratio]{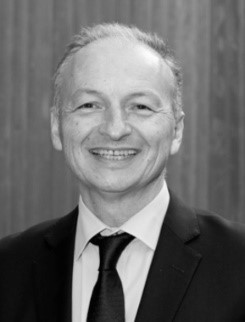}}]{Jovica V. Milanović} received the M.Sc. degree from the University of Belgrade, Yugoslavia, the Ph.D. degree from the University of Newcastle, Australia, and the D.Sc. degree from the University of Manchester U.K., all in electrical engineering. He is currently a Professor and Head of the Department of Electrical and Electronic Engineering at The University of Manchester, U.K. He is the Past Editor-in-Chief of IEEE Transactions on Power Systems, IEEE PES Governing Board member, Vice-Chair of the IEEE PES Fellows Evaluation Committee, Fellow of the Royal Academy of Engineering (UK), and IEEE and IET Fellow.
\end{IEEEbiography}

%
%\begin{IEEEbiography} [{\includegraphics[width=1in,height=1.25in,clip,keepaspectratio]{NING.png}}] {Ning Zhang} received the B.S. and Ph.D. degrees from Tsinghua University, Beijing, China, in 2007 and 2012, respectively. He is currently an Associate Professor at the same university. His research interests include power system planning, integrated energy systems, renewable energy integration, and low-carbon power system operation. He serves as an Associate Editor of the CSEE Journal of Power and Energy Systems and as an Editorial Board Member of Protection and Control of Modern Power Systems. 
%\end{IEEEbiography}
%\begin{IEEEbiography} [{\includegraphics[width=1in,height=1.25in,clip,keepaspectratio]{KANG.png}}] {Chongqing Kang} received the B.S. degree and the Ph.D. degree in Electrical Power Engineering from Tsinghua University, respectively in 1993 and 1997. He is currently a Professor, Dean of the Department of Electrical Engineering, and President of the Tsinghua University Energy Internet Research Institute. His research interests include power system planning and optimization, renewable energy integration, energy internet technologies, load forecasting, and electricity markets. 
%\end{IEEEbiography}


\begin{thebibliography}{99}

\bibitem{ref3}
N. Hatziargyriou \textit{et al.}, “Definition and classification of power system stability—Revisited and extended,” \textit{IEEE Trans. Power Syst.}, vol. 36, no. 4, pp. 3271–3281, Jul. 2021.
\bibitem{ref4}
Y. Dai, R. Preece, and M. Panteli, “Risk assessment of cascading failures in power systems with increasing wind penetration,” \textit{Electr. Power Syst. Res.}, vol. 211, p. 108392, 2022.

\bibitem{ref5}
H. Jia \textit{et al.}, “Learning multiple convex voltage stability constraints for unit commitment,” \textit{IEEE Trans. Power Syst.}, vol. 40, no. 1, pp. 125–137, Jan. 2025.

\bibitem{ref6}
E. Vittal, M. O’Malley, and A. Keane, “A steady-state voltage stability analysis of power systems with high penetrations of wind,” \textit{IEEE Trans. Power Syst.}, vol. 25, no. 1, pp. 433–442, Feb. 2010.

\bibitem{ref8}
A. Dissanayaka,\textit{et al.}, “Risk-based dynamic security assessment,” \textit{IEEE Trans. Power Syst.}, vol. 26, no. 3, pp. 1302–1308, Aug. 2011.

\bibitem{ref10}
M. Negnevitsky, D. H. Nguyen, and M. Piekutowski, “Risk assessment for power system operation planning with high wind power penetration,” \textit{IEEE Trans. Power Syst.}, vol. 30, no. 3, pp. 1359–1368, May 2015.

\bibitem{ref13}
J. A. Momoh, Y. V. Makarov, and W. Mittelstadt, “A framework of voltage stability assessment in power system reliability analysis,” \textit{IEEE Trans. Power Syst.}, vol. 14, no. 2, pp. 484–491, May 1999.

\bibitem{ref14}
K. N. Hasan, R. Preece, and J. V. Milanović, “Existing approaches and trends in uncertainty modelling and probabilistic stability analysis of power systems with renewable generation,” \textit{Renew. Sustain. Energy Rev.}, vol. 101, pp. 168–180, 2019.

\bibitem{ref15}
S. Liu, R. Sioshansi, and A. J. Conejo, “Hierarchical clustering to find representative operating periods for capacity-expansion modeling,” \textit{IEEE Trans. Power Syst.}, vol. 33, no. 3, pp. 3029–3039, May 2018.

\bibitem{ref16}
R. Alvarez, A. Moser, and C. A. Rahmann, “Novel methodology for selecting representative operating points for TNEP,” \textit{IEEE Trans. Power Syst.}, vol. 32, no. 3, pp. 2234–2242, May 2017.

\bibitem{ref17}
Q. Hou, E. Du, N. Zhang, and C. Kang, “Impact of high renewable penetration on the power system operation mode: A data-driven approach,” \textit{IEEE Trans. Power Syst.}, vol. 35, no. 1, pp. 731–741, Jan. 2020.

\bibitem{ref18}
K. Poncelet \textit{et al.}, “Selecting representative days for capturing the implications of integrating intermittent renewables in generation expansion planning problems,” \textit{IEEE Trans. Power Syst.}, vol. 32, no. 3, pp. 1936–1948, May 2017.

\bibitem{ref19}
S. Pineda and J. M. Morales, “Chronological time-period clustering for optimal capacity expansion planning with storage,” \textit{IEEE Trans. Power Syst.}, vol. 33, no. 6, pp. 7162–7170, Nov. 2018.

\bibitem{ref20}
G. Chicco, “Overview and performance assessment of the clustering methods for electrical load pattern grouping,” \textit{Energy}, vol. 42, no. 1, pp. 68–80, 2012.

\bibitem{ref22}
S. M. Miraftabzadeh \textit{et al.}, “K-means and alternative clustering methods in modern power systems,” \textit{IEEE Access}, vol. 11, pp. 119596–119633, 2023.

\bibitem{ref23}
CIGRE and IEEE, ‘Evaluation of Voltage Stability Assessment Methodologies in Modern Power Systems with High Penetration of Inverter-Based Resources - Technical Brochures’, JWG C4/C2.58/IEEE, 2024.  

\bibitem{ref24}
F. Murtagh and P. Legendre, “Ward’s hierarchical agglomerative clustering method: Which algorithms implement Ward’s criterion?” \textit{J. Classif.}, vol. 31, pp. 274–295, 2014.

\bibitem{ref25} 
Jin, X., Han, J. (2011). \textit{K}-Medoids Clustering. In: Sammut, C., Webb, G.I. (eds) Encyclopedia of Machine Learning. Springer, Boston, MA. 

\bibitem{ref026}
S. N. Singh, L. Srivastava, and J. Sharma, ‘Fast voltage contingency screening and ranking using cascade neural network’, Electric Power Systems Research, vol. 53, no. 3, pp. 197–205, Mar. 2000, doi: 10.1016/S0378-7796(99)00059-0.

\bibitem{ref027}
K. Nara et al., ‘On-Line Contingency Selection Algorithm for Voltage Security Analysis’, IEEE Transactions on Power Apparatus and Systems, vol. PAS-104, no. 4, pp. 846–856, Jul. 1985, doi: 10.1109/TPAS.1985.319085.


\bibitem{ref26}
T. Van Cutsem \textit{et al.}, “Test systems for voltage stability studies,” \textit{IEEE Trans. Power Syst.}, vol. 35, no. 5, pp. 4078–4087, Sept. 2020.

\bibitem{ref27}
Electric Power Research Institute, “The new aggregated distributed energy resources (DER\_A) model for transmission planning studies: 2019 update,” 2019.


\bibitem{ref33}
H. Cui \textit{et al}., "Disturbance Propagation in Power Grids With High Converter Penetration," in \textit{Proceedings of the IEEE}, vol. 111, no. 7, pp. 873-890, July 2023.

\bibitem{ref34}
M. Glavic and T. Van Cutsem, "Wide-Area Detection of Voltage Instability From Synchronized Phasor Measurements. Part II: Simulation Results," in \textit{IEEE Trans. Power Syst.}, vol. 24, no. 3, pp. 1417-1425, Aug. 2009 

\bibitem{ref35}
Y. Yang, Y. Sun, Q. Wang, F. Liu and L. Zhu, "Fast Power Grid Partition for Voltage Control With Balanced-Depth-Based Community Detection Algorithm," in \textit{IEEE Trans. Power Syst.}, vol. 37, no. 2, pp. 1612-1622, March 2022.

\bibitem{ref36}
A. D. Rajapakse et \textit{al.},, "Rotor Angle Instability Prediction Using Post-Disturbance Voltage Trajectories," in \textit{IEEE Trans. Power Syst.}, vol. 25, no. 2, pp. 947-956, May 2010. 

\end{thebibliography}
\end{document}